\documentclass[12pt,thmsa]{article}
\usepackage{sw20lart}

\input tcilatex
\begin{document}

\section{Introduction}

We have shown in several papers how the treatment of spin can be
generalized[1-5]. For spin 1/2[1-3] and spin 1[5], we have derived
generalized probability amplitudes for measurements on the systems, and we
have obtained generalized forms of the components of the spin operators. In
this paper, we extend this work to the case of spin 2. The method to be used
in obtaining the generalized quantities is given in full detail in refs 1-5.
Here the steps of the method will be used without description. The ideas
underlying the method owe their genesis to the work of Land\'e[6-9].

\section{Probability Amplitudes}

To obtain the generalized probability amplitudes, we start with the standard
expressions. These we derive by first obtaining the eigenvectors of the
operator that results from taking the dot product of the standard operator
and an arbitrary unit vector. Let the unit vector be $\widehat{\mathbf{a}},$
with polar angles $(\theta ,\varphi ).$ The standard operators for spin 2 are

\begin{equation}
\lbrack S_x]=\hbar \left( 
\begin{array}{ccccc}
0 & 1 & 0 & 0 & 0 \\ 
1 & 0 & \frac{\sqrt{6}}2 & 0 & 0 \\ 
0 & \frac{\sqrt{6}}2 & 0 & \frac{\sqrt{6}}2 & 0 \\ 
0 & 0 & \frac{\sqrt{6}}2 & 0 & 1 \\ 
0 & 0 & 0 & 1 & 0
\end{array}
\right)  \label{on1}
\end{equation}

\begin{equation}
\lbrack S_y]=\hbar \left( 
\begin{array}{ccccc}
0 & -i & 0 & 0 & 0 \\ 
i & 0 & -\frac{i\sqrt{6}}2 & 0 & 0 \\ 
0 & \frac{i\sqrt{6}}2 & 0 & -\frac{i\sqrt{6}}2 & 0 \\ 
0 & 0 & \frac{i\sqrt{6}}2 & 0 & -i \\ 
0 & 0 & 0 & i & 0
\end{array}
\right)  \label{tw2}
\end{equation}

and

\begin{equation}
\lbrack S_z]=\hbar \left( 
\begin{array}{ccccc}
2 & 0 & 0 & 0 & 0 \\ 
0 & 1 & 0 & 0 & 0 \\ 
0 & 0 & 0 & 0 & 0 \\ 
0 & 0 & 0 & -1 & 0 \\ 
0 & 0 & 0 & 0 & -2
\end{array}
\right)  \label{th3}
\end{equation}
so that $\mathbf{S}=\widehat{\mathbf{i}}S_x+\widehat{\mathbf{j}}S_y+\widehat{%
\mathbf{k}}S_z.$ Since $\widehat{\mathbf{a}}=(\sin \theta \cos \varphi ,\sin
\theta \sin \varphi ,\cos \theta )$, we have

\begin{equation}
\lbrack \widehat{\mathbf{a}}\cdot \mathbf{S]}=\hbar \left( 
\begin{array}{ccccc}
2\cos \theta & \sin \theta e^{-i\varphi } & 0 & 0 & 0 \\ 
\sin \theta e^{i\varphi } & \cos \theta & \frac{\sqrt{6}}2\sin \theta
e^{-i\varphi } & 0 & 0 \\ 
0 & \frac{\sqrt{6}}2\sin \theta e^{i\varphi } & 0 & \frac{\sqrt{6}}2\sin
\theta e^{-i\varphi } & 0 \\ 
0 & 0 & \frac{\sqrt{6}}2\sin \theta e^{i\varphi } & -\cos \theta & \sin
\theta e^{-i\varphi } \\ 
0 & 0 & 0 & \sin \theta e^{i\varphi } & -2\cos \theta
\end{array}
\right)  \label{fo4}
\end{equation}

The normalized eigenvectors of this matrix are

\begin{equation}
\lbrack \psi _2]=\left( 
\begin{array}{c}
\cos ^4\frac \theta 2e^{-i2\varphi } \\ 
\sin \theta \cos ^2\frac \theta 2e^{-i\varphi } \\ 
\frac{\sqrt{6}}4\sin ^2\theta \\ 
\sin \theta \sin ^2\frac \theta 2e^{i\varphi } \\ 
\sin ^4\frac \theta 2e^{i2\varphi }
\end{array}
\right)  \label{fi5}
\end{equation}

\begin{equation}
\lbrack \psi _1]=\left( 
\begin{array}{c}
\sin \theta \cos ^2\frac \theta 2e^{-i2\varphi } \\ 
(3\sin ^2\frac \theta 2-\cos ^2\frac \theta 2)\cos ^2\frac \theta
2e^{-i\varphi } \\ 
-\frac{\sqrt{6}}2\sin \theta \cos \theta \\ 
-(3\cos ^2\frac \theta 2-\sin ^2\frac \theta 2)\sin ^2\frac \theta
2e^{i\varphi } \\ 
-\sin \theta \sin ^2\frac \theta 2e^{i2\varphi }
\end{array}
\right)  \label{si6}
\end{equation}

\begin{equation}
\lbrack \psi _0]=\left( 
\begin{array}{c}
\frac{\sqrt{6}}4\sin ^2\theta e^{-i2\varphi } \\ 
-\frac{\sqrt{6}}2\sin \theta \cos \theta e^{-i\varphi } \\ 
\frac 12(2\cos ^2\theta -\sin ^2\theta ) \\ 
\frac{\sqrt{6}}2\sin \theta \cos \theta e^{i\varphi } \\ 
\frac{\sqrt{6}}4\sin ^2\theta e^{i2\varphi }
\end{array}
\right)  \label{se7}
\end{equation}
\begin{equation}
\lbrack \psi _{-1}]=\left( 
\begin{array}{c}
\sin \theta \sin ^2\frac \theta 2e^{-i2\varphi } \\ 
-(3\cos ^2\frac \theta 2-\sin ^2\frac \theta 2)\sin ^2\frac \theta
2e^{-i\varphi } \\ 
\frac{\sqrt{6}}2\sin \theta \cos \theta \\ 
(3\sin ^2\frac \theta 2-\cos ^2\frac \theta 2)\cos ^2\frac \theta
2e^{i\varphi } \\ 
-\sin \theta \cos ^2\frac \theta 2e^{i2\varphi }
\end{array}
\right)  \label{ei8}
\end{equation}
and

\begin{equation}
\lbrack \psi _{-2}]=\left( 
\begin{array}{c}
\sin ^4\frac \theta 2e^{-i2\varphi } \\ 
-\sin \theta \sin ^2\frac \theta 2e^{-i\varphi } \\ 
\frac{\sqrt{6}}4\sin ^2\theta \\ 
-\sin \theta \cos ^2\frac \theta 2e^{i\varphi } \\ 
\cos ^4\frac \theta 2e^{i2\varphi }
\end{array}
\right)  \label{ni9}
\end{equation}

These vectors correspond to the case where the initial quantization
direction is the vector $\widehat{\mathbf{a}}$ (whose polar angles are $%
(\theta ,\varphi )$), while the final quantization direction is the $z$
axis. As has been pointed out, the elements of this eigenvector are
probability amplitudes of the form $\psi (m_i^{(\widehat{\mathbf{a}})};m_f^{(%
\widehat{\mathbf{k}})})$[1]$.$ The square modulus of the probability
amplitude $\psi (m_i^{(\widehat{\mathbf{a}})};m_f^{(\widehat{\mathbf{k}})})$
gives the probability that if the system is initially in a state of spin
projection $m_i\hbar $ in the direction $\widehat{\mathbf{a}}$, a
measurement of the projection in the direction $\widehat{\mathbf{k}}$ yields
the value $m_f\hbar $. Hence, we are able to deduce that

\begin{equation}
\psi (2^{(\widehat{\mathbf{a}})};2^{(\widehat{\mathbf{k}})})=\cos ^4\frac
\theta 2e^{-i2\varphi },  \label{te10}
\end{equation}
\begin{equation}
\psi (2^{(\widehat{\mathbf{a}})};1^{(\widehat{\mathbf{k}})})=\sin \theta
\cos ^2\frac \theta 2e^{-i\varphi },  \label{ei11}
\end{equation}
\begin{equation}
\psi (2^{(\widehat{\mathbf{a}})};0^{(\widehat{\mathbf{k}})})=\frac{\sqrt{6}}%
4\sin ^2\theta ,  \label{tw12}
\end{equation}
\begin{equation}
\psi (2^{(\widehat{\mathbf{a}})};(-1)^{(\widehat{\mathbf{k}})})=\sin \theta
\sin ^2\frac \theta 2e^{i\varphi },  \label{th13}
\end{equation}
\begin{equation}
\psi (2^{(\widehat{\mathbf{a}})};(-2)^{(\widehat{\mathbf{k}})})=\sin ^4\frac
\theta 2e^{i2\varphi },  \label{fo14}
\end{equation}

\begin{equation}
\psi (1^{(\widehat{\mathbf{a}})};2^{(\widehat{\mathbf{k}})})=\sin \theta
\cos ^2\frac \theta 2e^{-i2\varphi },  \label{fi15}
\end{equation}

\begin{equation}
\psi (1^{(\widehat{\mathbf{a}})};1^{(\widehat{\mathbf{k}})})=(3\sin ^2\frac
\theta 2-\cos ^2\frac \theta 2)\cos ^2\frac \theta 2e^{-i\varphi },
\label{si16}
\end{equation}
\begin{equation}
\psi (1^{(\widehat{\mathbf{a}})};0^{(\widehat{\mathbf{k}})})=-\frac{\sqrt{6}}%
2\sin \theta \cos \theta ,  \label{se17}
\end{equation}
\begin{equation}
\psi (1^{(\widehat{\mathbf{a}})};(-1)^{(\widehat{\mathbf{k}})})=-(3\cos
^2\frac \theta 2-\sin ^2\frac \theta 2)\sin ^2\frac \theta 2e^{i\varphi },
\label{ei18}
\end{equation}

\begin{equation}
\psi (1^{(\widehat{\mathbf{a}})};(-2)^{(\widehat{\mathbf{k}})})=-\sin \theta
\sin ^2\frac \theta 2e^{i2\varphi },  \label{ni19}
\end{equation}
\begin{equation}
\psi (0^{(\widehat{\mathbf{a}})};2^{(\widehat{\mathbf{k}})})=\frac{\sqrt{6}}%
4\sin ^2\theta e^{-i2\varphi },  \label{tw20}
\end{equation}
\begin{equation}
\psi (0^{(\widehat{\mathbf{a}})};1^{(\widehat{\mathbf{k}})})=-\frac{\sqrt{6}}%
2\sin \theta \cos \theta e^{-i\varphi },  \label{tw21}
\end{equation}

\begin{equation}
\psi (0^{(\widehat{\mathbf{a}})};0^{(\widehat{\mathbf{k}})})=\frac 12(2\cos
^2\theta -\sin ^2\theta ),  \label{tw22}
\end{equation}
\begin{equation}
\psi (0^{(\widehat{\mathbf{a}})};(-1)^{(\widehat{\mathbf{k}})})=\frac{\sqrt{6%
}}2\sin \theta \cos \theta e^{i\varphi },  \label{tw23}
\end{equation}
\begin{equation}
\psi (0^{(\widehat{\mathbf{a}})};(-2)^{(\widehat{\mathbf{k}})})=\frac{\sqrt{6%
}}4\sin ^2\theta e^{i2\varphi },  \label{tw24}
\end{equation}

\begin{equation}
\psi ((-1)^{(\widehat{\mathbf{a}})};2^{(\widehat{\mathbf{k}})})=\sin \theta
\sin ^2\frac \theta 2e^{-i2\varphi },  \label{tw25}
\end{equation}
\begin{equation}
\psi ((-1)^{(\widehat{\mathbf{a}})};1^{(\widehat{\mathbf{k}})})=-(3\cos
^2\frac \theta 2-\sin ^2\frac \theta 2)\sin ^2\frac \theta 2e^{-i\varphi },
\label{tw26}
\end{equation}
\begin{equation}
\psi ((-1)^{(\widehat{\mathbf{a}})};0^{(\widehat{\mathbf{k}})})=\frac{\sqrt{6%
}}2\sin \theta \cos \theta ,  \label{tw27}
\end{equation}
\begin{equation}
\psi ((-1)^{(\widehat{\mathbf{a}})};(-1)^{(\widehat{\mathbf{k}})})=(3\sin
^2\frac \theta 2-\cos ^2\frac \theta 2)\cos ^2\frac \theta 2e^{i\varphi },
\label{tw28}
\end{equation}
\begin{equation}
\psi ((-1)^{(\widehat{\mathbf{a}})};(-2)^{(\widehat{\mathbf{k}})})=-\sin
\theta \cos ^2\frac \theta 2e^{i2\varphi },  \label{tw29}
\end{equation}

\begin{equation}
\psi ((-2)^{(\widehat{\mathbf{a}})};2^{(\widehat{\mathbf{k}})})=\sin ^4\frac
\theta 2e^{-i2\varphi },  \label{th30}
\end{equation}
\begin{equation}
\psi ((-2)^{(\widehat{\mathbf{a}})};1^{(\widehat{\mathbf{k}})})=-\sin \theta
\sin ^2\frac \theta 2e^{-i\varphi },  \label{th31}
\end{equation}
\begin{equation}
\psi ((-2)^{(\widehat{\mathbf{a}})};0^{(\widehat{\mathbf{k}})})=\frac{\sqrt{6%
}}4\sin ^2\theta ,  \label{th32}
\end{equation}
\begin{equation}
\psi ((-2)^{(\widehat{\mathbf{a}})};(-1)^{(\widehat{\mathbf{k}})})=-\sin
\theta \cos ^2\frac \theta 2e^{i\varphi }  \label{th33}
\end{equation}
and 
\begin{equation}
\psi ((-2)^{(\widehat{\mathbf{a}})};(-2)^{(\widehat{\mathbf{k}})})=\cos
^4\frac \theta 2e^{i2\varphi }.  \label{th34}
\end{equation}

\section{Generalized Probability Amplitudes}

We now consider the general case. Here, the original direction of
quantization is given by the vector $\widehat{\mathbf{a}}(\theta ^{\prime
},\varphi ^{\prime })$ and the final direction by the vector $\widehat{%
\mathbf{c}}(\theta ,\varphi ).$ Then the eigenvectors are given by

\begin{equation}
\lbrack \psi _{m_i}]=\left( 
\begin{array}{c}
\psi (m_i^{(\widehat{\mathbf{a}})};2^{(\widehat{\mathbf{c}})}) \\ 
\psi (m_i^{(\widehat{\mathbf{a}})};1^{(\widehat{\mathbf{c}})}) \\ 
\psi (m_i^{(\widehat{\mathbf{a}})};0^{(\widehat{\mathbf{c}})}) \\ 
\psi (m_i^{(\widehat{\mathbf{a}})};(-1)^{(\widehat{\mathbf{c}})}) \\ 
\psi (m_i^{(\widehat{\mathbf{a}})};(-2)^{(\widehat{\mathbf{c}})})
\end{array}
\right)  \label{th35}
\end{equation}
To obtain the generalized probability amplitudes $\psi (m_i^{(\widehat{%
\mathbf{a}})};m_f^{(\widehat{\mathbf{c}})}),$ we use the expansion[6] 
\begin{equation}
\psi (m_i^{(\widehat{\mathbf{a}})};m_f^{(\widehat{\mathbf{c}}%
)})=\sum_{m=-2}^2\psi (m_i^{(\widehat{\mathbf{a}})};m^{(\widehat{\mathbf{b}}%
)})\psi (m^{(\widehat{\mathbf{b}})};m_f^{(\widehat{\mathbf{c}})})
\label{th36}
\end{equation}
where $\widehat{\mathbf{b}}$ is an intermediate arbitrary direction, whose
polar angles never appear in the final expressions. This is a particular
form of the expansion

\begin{equation}
\psi (A_i,C_k)=\dsum\limits_{j=1}^N\chi (A_i,B_j)\phi (B_j,C_k).
\label{th37}
\end{equation}
which is always true for any three observables $A$, $B$ and $C$ of a system.
Here $\psi $ are the probability amplitudes for measurement of $C$ from the
state in which the system is initially in an eigenstate of $A$; $\chi $ are
the probability amplitudes for measurement of $B$ for the same case. In the
same way, $\phi $ are the probability amplitudes for measurement of $C$ when
the system is initially in an eigenstate of $B$. Here, $A_i$, $B_i$ and $C_i$
are eigenvalues of the observables $A$, $B$ and $C.$

For the present case, the eigenvector for the eigenvalue $2\hbar $ is

\begin{equation}
\lbrack \psi _2]=\left( 
\begin{array}{c}
\psi (2^{(\widehat{\mathbf{a}})};2^{(\widehat{\mathbf{c}})}) \\ 
\psi (2^{(\widehat{\mathbf{a}})};1^{(\widehat{\mathbf{c}})}) \\ 
\psi (2^{(\widehat{\mathbf{a}})};0^{(\widehat{\mathbf{c}})}) \\ 
\psi (2^{(\widehat{\mathbf{a}})};(-1)^{(\widehat{\mathbf{c}})}) \\ 
\psi (2^{(\widehat{\mathbf{a}})};(-2)^{(\widehat{\mathbf{c}})})
\end{array}
\right)  \label{th38}
\end{equation}
With the aid of Eqs. (\ref{te10}) - (\ref{th34}) and the expansion Eq. (\ref
{th36}), we find that the generalized probability amplitudes are

\begin{eqnarray}
\ &&\psi (2^{(\widehat{\mathbf{a}})};2^{(\widehat{\mathbf{c}})})=\cos
^4\frac \theta 2\cos ^4\frac{\theta ^{\prime }}2e^{i2(\varphi -\varphi
^{\prime })}  \nonumber \\
&&\ +\sin \theta ^{\prime }\sin \theta \cos ^2\frac{\theta ^{\prime }}2\cos
^2\frac \theta 2e^{i(\varphi -\varphi ^{\prime })}+\frac 38\sin ^2\theta
^{\prime }\sin ^2\theta  \nonumber \\
&&\ +\sin \theta ^{\prime }\sin \theta \sin ^2\frac{\theta ^{\prime }}2\sin
^2\frac \theta 2e^{-i(\varphi -\varphi ^{\prime })}  \nonumber \\
&&\ +\sin ^4\frac \theta 2\sin ^4\frac{\theta ^{\prime }}2e^{-i2(\varphi
-\varphi ^{\prime })},  \label{th39}
\end{eqnarray}

\begin{eqnarray}
\ &&\psi (2^{(\widehat{\mathbf{a}})};1^{(\widehat{\mathbf{c}})})=\sin \theta
\cos ^4\frac{\theta ^{\prime }}2\cos ^2\frac \theta 2e^{i2(\varphi -\varphi
^{\prime })}  \nonumber \\
&&\ +(3\sin ^2\frac \theta 2-\cos ^2\frac \theta 2)\sin \theta ^{\prime
}\cos ^2\frac{\theta ^{\prime }}2\cos ^2\frac \theta 2e^{i(\varphi -\varphi
^{\prime })}  \nonumber \\
&&\ -\frac 34\sin ^2\theta ^{\prime }\sin \theta \cos \theta  \nonumber \\
&&\ -(3\cos ^2\frac \theta 2-\sin ^2\frac \theta 2)\sin \theta ^{\prime
}\sin ^2\frac{\theta ^{\prime }}2\sin ^2\frac \theta 2e^{-i(\varphi -\varphi
^{\prime })}  \nonumber \\
&&-\sin \theta \sin ^4\frac{\theta ^{\prime }}2\sin ^2\frac \theta
2e^{-i2(\varphi -\varphi ^{\prime })}  \label{fo40}
\end{eqnarray}

\begin{eqnarray}
\ &&\ \psi (2^{(\widehat{\mathbf{a}})};0^{(\widehat{\mathbf{c}})})=\sqrt{%
\frac 38}\sin ^2\theta \cos ^4\frac{\theta ^{\prime }}2e^{i2(\varphi
-\varphi ^{\prime })}  \nonumber \\
&&\ -\sqrt{\frac 32}\sin \theta \sin \theta ^{\prime }\cos \theta \cos ^2%
\frac{\theta ^{\prime }}2e^{i(\varphi -\varphi ^{\prime })}+\sqrt{\frac 3{32}%
}(2\cos ^2\theta -\sin ^2\theta )\sin ^2\theta ^{\prime }  \nonumber \\
&&\ +\sqrt{\frac 32}\sin \theta \sin \theta ^{\prime }\cos \theta \sin ^2%
\frac{\theta ^{\prime }}2e^{-i(\varphi -\varphi ^{\prime })}+\sqrt{\frac 38}%
\sin ^2\theta \sin ^4\frac{\theta ^{\prime }}2e^{-i2(\varphi -\varphi
^{\prime })}  \nonumber \\
&&  \label{fo41}
\end{eqnarray}
\begin{eqnarray}
\ &&\psi (2^{(\widehat{\mathbf{a}})};(-1)^{(\widehat{\mathbf{c}})})=\sin
\theta \cos ^4\frac{\theta ^{\prime }}2\sin ^2\frac \theta 2e^{i2(\varphi
-\varphi ^{\prime })}  \nonumber \\
&&\ \ -(3\cos ^2\frac \theta 2-\sin ^2\frac \theta 2)\sin \theta ^{\prime
}\cos ^2\frac{\theta ^{\prime }}2\sin ^2\frac \theta 2e^{i(\varphi -\varphi
^{\prime })}  \nonumber \\
&&\ \ +\frac 34\sin ^2\theta ^{\prime }\sin \theta \cos \theta  \nonumber \\
&&\ \ +(3\sin ^2\frac \theta 2-\cos ^2\frac \theta 2)\sin \theta ^{\prime
}\sin ^2\frac{\theta ^{\prime }}2\cos ^2\frac \theta 2e^{-i(\varphi -\varphi
^{\prime })}  \nonumber \\
&&-\sin \theta \sin ^4\frac{\theta ^{\prime }}2\cos ^2\frac \theta
2e^{-i2(\varphi -\varphi ^{\prime })}  \label{fo42}
\end{eqnarray}
and 
\begin{eqnarray}
\ &&\psi (2^{(\widehat{\mathbf{a}})};(-2)^{(\widehat{\mathbf{c}})})=\sin
^4\frac \theta 2\cos ^4\frac{\theta ^{\prime }}2e^{i2(\varphi -\varphi
^{\prime })}  \nonumber \\
&&\ \ -\sin \theta ^{\prime }\sin \theta \cos ^2\frac{\theta ^{\prime }}%
2\sin ^2\frac \theta 2e^{i(\varphi -\varphi ^{\prime })}+\frac 38\sin
^2\theta ^{\prime }\sin ^2\theta  \nonumber \\
&&\ \ -\sin \theta ^{\prime }\sin \theta \sin ^2\frac{\theta ^{\prime }}%
2\cos ^2\frac \theta 2e^{-i(\varphi -\varphi ^{\prime })}  \nonumber \\
&&\ \ +\cos ^4\frac \theta 2\sin ^4\frac{\theta ^{\prime }}2e^{-i2(\varphi
-\varphi ^{\prime })}.  \label{fo43}
\end{eqnarray}

For the eigenvalue $\hbar $, the vector is

\begin{equation}
\lbrack \psi _1]=\left( 
\begin{array}{c}
\psi (1^{(\widehat{\mathbf{a}})};2^{(\widehat{\mathbf{c}})}) \\ 
\psi (1^{(\widehat{\mathbf{a}})};1^{(\widehat{\mathbf{c}})}) \\ 
\psi (1^{(\widehat{\mathbf{a}})};0^{(\widehat{\mathbf{c}})}) \\ 
\psi (1^{(\widehat{\mathbf{a}})};(-1)^{(\widehat{\mathbf{c}})}) \\ 
\psi (1^{(\widehat{\mathbf{a}})};(-2)^{(\widehat{\mathbf{c}})})
\end{array}
\right)  \label{fo44}
\end{equation}
where the probability amplitudes are 
\begin{eqnarray}
\ &&\psi (1^{(\widehat{\mathbf{a}})};2^{(\widehat{\mathbf{c}})})=\sin \theta
^{\prime }\cos ^2\frac{\theta ^{\prime }}2\cos ^4\frac \theta 2e^{i2(\varphi
-\varphi ^{\prime })}  \nonumber \\
&&\ \ \ +(3\sin ^2\frac{\theta ^{\prime }}2-\cos ^2\frac{\theta ^{\prime }}%
2)\sin \theta \cos ^2\frac{\theta ^{\prime }}2\cos ^2\frac \theta
2e^{i(\varphi -\varphi ^{\prime })}  \nonumber \\
&&\ \ \ -\frac 34\sin ^2\theta \sin \theta ^{\prime }\cos \theta ^{\prime } 
\nonumber \\
&&\ \ \ +(3\cos ^2\frac{\theta ^{\prime }}2-\sin ^2\frac{\theta ^{\prime }}%
2)\sin \theta \sin ^2\frac{\theta ^{\prime }}2\sin ^2\frac \theta
2e^{-i(\varphi -\varphi ^{\prime })}  \nonumber \\
&&-\sin \theta ^{\prime }\sin ^2\frac{\theta ^{\prime }}2\sin ^4\frac \theta
2e^{-i2(\varphi -\varphi ^{\prime })}  \label{fo45}
\end{eqnarray}

\begin{eqnarray}
\ &&\psi (1^{(\widehat{\mathbf{a}})};1^{(\widehat{\mathbf{c}})})=\sin \theta
^{\prime }\sin \theta \cos ^2\frac{\theta ^{\prime }}2\cos ^2\frac \theta
2e^{i2(\varphi -\varphi ^{\prime })}  \nonumber \\
&&\ \ \ \ +(3\sin ^2\frac{\theta ^{\prime }}2-\cos ^2\frac{\theta ^{\prime }}%
2)(3\sin ^2\frac \theta 2-\cos ^2\frac \theta 2)\cos ^2\frac{\theta ^{\prime
}}2\cos ^2\frac \theta 2e^{i(\varphi -\varphi ^{\prime })}  \nonumber \\
&&+\frac 32\sin \theta \sin \theta ^{\prime }\cos \theta \cos \theta
^{\prime }  \nonumber \\
&&\ \ \ \ +(3\cos ^2\frac{\theta ^{\prime }}2-\sin ^2\frac{\theta ^{\prime }}%
2)(3\cos ^2\frac \theta 2-\sin ^2\frac \theta 2)\sin ^2\frac{\theta ^{\prime
}}2\sin ^2\frac \theta 2e^{-i(\varphi -\varphi ^{\prime })}  \nonumber \\
&&\ \ \ \ +\sin \theta ^{\prime }\sin \theta \sin ^2\frac{\theta ^{\prime }}%
2\sin ^2\frac \theta 2e^{-i2(\varphi -\varphi ^{\prime })},  \label{fo46}
\end{eqnarray}

\begin{eqnarray}
\ &&\psi (1^{(\widehat{\mathbf{a}})};0^{(\widehat{\mathbf{c}})})=\sqrt{\frac
38}\sin \theta ^{\prime }\sin ^2\theta \cos ^2\frac{\theta ^{\prime }}%
2e^{i2(\varphi -\varphi ^{\prime })}  \nonumber \\
&&\ \ \ \ \ \ -\sqrt{\frac 32}(3\sin ^2\frac{\theta ^{\prime }}2-\cos ^2%
\frac{\theta ^{\prime }}2)\sin \theta \cos \theta \cos ^2\frac{\theta
^{\prime }}2e^{i(\varphi -\varphi ^{\prime })}  \nonumber \\
&&-\sqrt{\frac 38}(2\cos ^2\theta -\sin ^2\theta )\sin \theta ^{\prime }\cos
\theta ^{\prime }  \nonumber \\
&&\ \ \ \ \ \ -\sqrt{\frac 32}(3\cos ^2\frac{\theta ^{\prime }}2-\sin ^2%
\frac{\theta ^{\prime }}2)\sin \theta \cos \theta \sin ^2\frac{\theta
^{\prime }}2e^{-i(\varphi -\varphi ^{\prime })}  \nonumber \\
&&\ -\sqrt{\frac 38}\sin \theta ^{\prime }\sin ^2\theta \sin ^2\frac{\theta
^{\prime }}2e^{-i2(\varphi -\varphi ^{\prime })},  \label{fo47}
\end{eqnarray}
and

\begin{eqnarray}
\ &&\psi (1^{(\widehat{\mathbf{a}})};(-1)^{(\widehat{\mathbf{c}})})=\sin
\theta ^{\prime }\sin \theta \cos ^2\frac{\theta ^{\prime }}2\sin ^2\frac
\theta 2e^{i2(\varphi -\varphi ^{\prime })}  \nonumber \\
&&\ \ \ \ \ -(3\sin ^2\frac{\theta ^{\prime }}2-\cos ^2\frac{\theta ^{\prime
}}2)(3\cos ^2\frac \theta 2-\sin ^2\frac \theta 2)\cos ^2\frac{\theta
^{\prime }}2\sin ^2\frac \theta 2e^{i(\varphi -\varphi ^{\prime })} 
\nonumber \\
&&-\frac 32\sin \theta \sin \theta ^{\prime }\cos \theta \cos \theta
^{\prime }  \nonumber \\
&&\ \ \ \ \ -(3\cos ^2\frac{\theta ^{\prime }}2-\sin ^2\frac{\theta ^{\prime
}}2)(3\sin ^2\frac \theta 2-\cos ^2\frac \theta 2)\sin ^2\frac{\theta
^{\prime }}2\cos ^2\frac \theta 2e^{-i(\varphi -\varphi ^{\prime })} 
\nonumber \\
&&\ \ \ \ \ +\sin \theta ^{\prime }\sin \theta \sin ^2\frac{\theta ^{\prime }%
}2\cos ^2\frac \theta 2e^{-i2(\varphi -\varphi ^{\prime })},  \label{fo48}
\end{eqnarray}
\begin{eqnarray}
\ &&\psi (1^{(\widehat{\mathbf{a}})};(-2)^{(\widehat{\mathbf{c}})})=\sin
\theta ^{\prime }\cos ^2\frac{\theta ^{\prime }}2\sin ^4\frac \theta
2e^{i2(\varphi -\varphi ^{\prime })}  \nonumber \\
&&\ \ \ \ -(3\sin ^2\frac{\theta ^{\prime }}2-\cos ^2\frac{\theta ^{\prime }}%
2)\sin \theta \cos ^2\frac{\theta ^{\prime }}2\sin ^2\frac \theta
2e^{i(\varphi -\varphi ^{\prime })}  \nonumber \\
&&\ \ \ \ -\frac 34\sin ^2\theta \sin \theta ^{\prime }\cos \theta ^{\prime
}-\sin \theta ^{\prime }\sin ^2\frac{\theta ^{\prime }}2\cos ^4\frac \theta
2e^{-i2(\varphi -\varphi ^{\prime })}  \nonumber \\
&&\ \ \ \ +(3\cos ^2\frac{\theta ^{\prime }}2-\sin ^2\frac{\theta ^{\prime }}%
2)\sin \theta \sin ^2\frac{\theta ^{\prime }}2\cos ^2\frac \theta
2e^{-i(\varphi -\varphi ^{\prime })},  \label{fo49}
\end{eqnarray}
For the eigenvalue $0$, the vector is

\begin{equation}
\lbrack \psi _0]=\left( 
\begin{array}{c}
\psi (0^{(\widehat{\mathbf{a}})};2^{(\widehat{\mathbf{c}})}) \\ 
\psi (0^{(\widehat{\mathbf{a}})};1^{(\widehat{\mathbf{c}})}) \\ 
\psi (0^{(\widehat{\mathbf{a}})};0^{(\widehat{\mathbf{c}})}) \\ 
\psi (0^{(\widehat{\mathbf{a}})};(-1)^{(\widehat{\mathbf{c}})}) \\ 
\psi (0^{(\widehat{\mathbf{a}})};(-2)^{(\widehat{\mathbf{c}})})
\end{array}
\right)  \label{fi50}
\end{equation}
with the probability amplitudes 
\begin{eqnarray}
\ &&\ \psi (0^{(\widehat{\mathbf{a}})};2^{(\widehat{\mathbf{c}})})=\sqrt{%
\frac 38}\sin ^2\theta ^{\prime }\cos ^4\frac \theta 2e^{i2(\varphi -\varphi
^{\prime })}  \nonumber \\
&&\ \ -\sqrt{\frac 32}\sin \theta \sin \theta ^{\prime }\cos \theta ^{\prime
}\cos ^2\frac \theta 2e^{i(\varphi -\varphi ^{\prime })}  \nonumber \\
&&\ \ +\sqrt{\frac 3{32}}(2\cos ^2\theta ^{\prime }-\sin ^2\theta ^{\prime
})\sin ^2\theta  \nonumber \\
&&\ \ +\sqrt{\frac 32}\sin \theta \sin \theta ^{\prime }\cos \theta ^{\prime
}\sin ^2\frac \theta 2e^{-i(\varphi -\varphi ^{\prime })}  \nonumber \\
&&+\sqrt{\frac 38}\sin ^2\theta ^{\prime }\sin ^4\frac \theta
2e^{-i2(\varphi -\varphi ^{\prime })}  \label{fi51}
\end{eqnarray}

\begin{center}
\begin{eqnarray}
\ &&\psi (0^{(\widehat{\mathbf{a}})};1^{(\widehat{\mathbf{c}})})=\sqrt{\frac
38}\sin \theta \sin ^2\theta ^{\prime }\cos ^2\frac \theta 2e^{i2(\varphi
-\varphi ^{\prime })}  \nonumber \\
&&\ \ \ \ \ \ -\sqrt{\frac 32}(3\sin ^2\frac \theta 2-\cos ^2\frac \theta
2)\sin \theta ^{\prime }\cos \theta ^{\prime }\cos ^2\frac \theta
2e^{i(\varphi -\varphi ^{\prime })}  \nonumber \\
&&-\sqrt{\frac 38}(2\cos ^2\theta ^{\prime }-\sin ^2\theta ^{\prime })\sin
\theta \cos \theta  \nonumber \\
&&\ \ \ \ \ \ -\sqrt{\frac 32}(3\cos ^2\frac \theta 2-\sin ^2\frac \theta
2)\sin \theta ^{\prime }\cos \theta ^{\prime }\sin ^2\frac \theta
2e^{-i(\varphi -\varphi ^{\prime })}  \nonumber \\
&&\ -\sqrt{\frac 38}\sin \theta \sin ^2\theta ^{\prime }\sin ^2\frac \theta
2e^{-i2(\varphi -\varphi ^{\prime })},  \label{fi52}
\end{eqnarray}
\begin{eqnarray}
\ &&\psi (0^{(\widehat{\mathbf{a}})};0^{(\widehat{\mathbf{c}})})=\frac
38\sin ^2\theta \sin ^2\theta ^{\prime }e^{i2(\varphi -\varphi ^{\prime })} 
\nonumber \\
&&\ \ \ \ \ \ \ +\frac 32\sin \theta ^{\prime }\cos \theta ^{\prime }\sin
\theta \cos \theta e^{i(\varphi -\varphi ^{\prime })}+  \nonumber \\
&&\ \ \ \ \ \ \ +\frac 32\sin \theta ^{\prime }\cos \theta ^{\prime }\sin
\theta \cos \theta e^{-i(\varphi -\varphi ^{\prime })}  \nonumber \\
&&\ \ \ \ \ \ \ +\frac 14(2\cos ^2\theta ^{\prime }-\sin ^2\theta ^{\prime
})(2\cos ^2\theta -\sin ^2\theta )  \nonumber \\
&&\ \ +\frac 38\sin ^2\theta \sin ^2\theta ^{\prime }e^{-i2(\varphi -\varphi
^{\prime })},  \label{fi53}
\end{eqnarray}
\end{center}

and

\begin{eqnarray}
\ &&\psi (0^{(\widehat{\mathbf{a}})};(-1)^{(\widehat{\mathbf{c}})})=\sqrt{%
\frac 38}\sin \theta \sin ^2\theta ^{\prime }\sin ^2\frac \theta
2e^{i2(\varphi -\varphi ^{\prime })}  \nonumber \\
&&\ \ \ \ \ \ \ +\sqrt{\frac 32}(3\cos ^2\frac \theta 2-\sin ^2\frac \theta
2)\sin \theta ^{\prime }\cos \theta ^{\prime }\sin ^2\frac \theta
2e^{i(\varphi -\varphi ^{\prime })}  \nonumber \\
&&\ \ \ \ \ \ \ +\sqrt{\frac 32}(3\sin ^2\frac \theta 2-\cos ^2\frac \theta
2)\sin \theta ^{\prime }\cos \theta ^{\prime }\cos ^2\frac \theta
2e^{-i(\varphi -\varphi ^{\prime })}  \nonumber \\
&&\ \ \ \ \ \ \ +\sqrt{\frac 38}(2\cos ^2\theta ^{\prime }-\sin ^2\theta
^{\prime })\sin \theta \cos \theta  \nonumber \\
&&\ \ -\sqrt{\frac 38}\sin \theta \sin ^2\theta ^{\prime }\cos ^2\frac
\theta 2e^{-i2(\varphi -\varphi ^{\prime })},  \label{fi54}
\end{eqnarray}
\begin{eqnarray}
\ &&\ \psi (0^{(\widehat{\mathbf{a}})};(-2)^{(\widehat{\mathbf{c}})})=\sqrt{%
\frac 38}\sin ^2\theta ^{\prime }\sin ^4\frac \theta 2e^{i2(\varphi -\varphi
^{\prime })}  \nonumber \\
&&\ \ \ \ +\sqrt{\frac 32}\sin \theta \sin \theta ^{\prime }\cos \theta
^{\prime }\sin ^2\frac \theta 2e^{i(\varphi -\varphi ^{\prime })}+\sqrt{%
\frac 38}\sin ^2\theta ^{\prime }\cos ^4\frac \theta 2e^{-i2(\varphi
-\varphi ^{\prime })}  \nonumber \\
&&\ \ \ \ +\sqrt{\frac 3{32}}(2\cos ^2\theta ^{\prime }-\sin ^2\theta
^{\prime })\sin ^2\theta  \nonumber \\
&&\ \ \ \ +\sqrt{\frac 32}\sin \theta \sin \theta ^{\prime }\cos \theta
^{\prime }\cos ^2\frac \theta 2e^{-i(\varphi -\varphi ^{\prime })},
\label{fi55}
\end{eqnarray}

Finally, for the eigenvalue $-\hbar $, the vector is

\begin{equation}
\lbrack \psi _{-1}]=\left( 
\begin{array}{c}
\psi ((-1)^{(\widehat{\mathbf{a}})};2^{(\widehat{\mathbf{c}})}) \\ 
\psi ((-1)^{(\widehat{\mathbf{a}})};1^{(\widehat{\mathbf{c}})}) \\ 
\psi ((-1)^{(\widehat{\mathbf{a}})};0^{(\widehat{\mathbf{c}})}) \\ 
\psi ((-1)^{(\widehat{\mathbf{a}})};(-1)^{(\widehat{\mathbf{c}})}) \\ 
\psi ((-1)^{(\widehat{\mathbf{a}})};(-2)^{(\widehat{\mathbf{c}})})
\end{array}
\right)  \label{fi56}
\end{equation}
where the probability amplitudes are

\begin{eqnarray}
\ &&\psi ((-1)^{(\widehat{\mathbf{a}})};2^{(\widehat{\mathbf{c}})})=-\sin
\theta ^{\prime }\sin ^2\frac{\theta ^{\prime }}2\cos ^4\frac \theta
2e^{i2(\varphi -\varphi ^{\prime })}  \nonumber \\
&&\ \ \ \ \ -(3\cos ^2\frac{\theta ^{\prime }}2-\sin ^2\frac{\theta ^{\prime
}}2)\sin \theta \sin ^2\frac{\theta ^{\prime }}2\cos ^2\frac \theta
2e^{i(\varphi -\varphi ^{\prime })}  \nonumber \\
&&\ \ \ \ \ +\frac 34\sin ^2\theta \sin \theta ^{\prime }\cos \theta
^{\prime }-\sin \theta ^{\prime }\cos ^2\frac{\theta ^{\prime }}2\sin
^4\frac \theta 2e^{-i2(\varphi -\varphi ^{\prime })}  \nonumber \\
&&\ \ \ \ \ +(3\sin ^2\frac{\theta ^{\prime }}2-\cos ^2\frac{\theta ^{\prime
}}2)\sin \theta \cos ^2\frac{\theta ^{\prime }}2\sin ^2\frac \theta
2e^{-i(\varphi -\varphi ^{\prime })},  \label{fi57}
\end{eqnarray}

\begin{eqnarray}
\ &&\psi ((-1)^{(\widehat{\mathbf{a}})};1^{(\widehat{\mathbf{c}})})=\sin
\theta ^{\prime }\sin \theta \sin ^2\frac{\theta ^{\prime }}2\cos ^2\frac
\theta 2e^{i2(\varphi -\varphi ^{\prime })}  \nonumber \\
&&\ \ \ \ \ \ \ -(3\cos ^2\frac{\theta ^{\prime }}2-\sin ^2\frac{\theta
^{\prime }}2)(3\sin ^2\frac \theta 2-\cos ^2\frac \theta 2)\sin ^2\frac{%
\theta ^{\prime }}2\cos ^2\frac \theta 2e^{i(\varphi -\varphi ^{\prime })} 
\nonumber \\
&&\ \ \ \ \ \ \ -(3\sin ^2\frac{\theta ^{\prime }}2-\cos ^2\frac{\theta
^{\prime }}2)(3\cos ^2\frac \theta 2-\sin ^2\frac \theta 2)\cos ^2\frac{%
\theta ^{\prime }}2\sin ^2\frac \theta 2e^{-i(\varphi -\varphi ^{\prime })} 
\nonumber \\
&&\ \ \ \ \ \ \ -\frac 32\sin \theta \sin \theta ^{\prime }\cos \theta \cos
\theta ^{\prime }+\sin \theta ^{\prime }\sin \theta \cos ^2\frac{\theta
^{\prime }}2\sin ^2\frac \theta 2e^{-i2(\varphi -\varphi ^{\prime })},
\label{fi58}
\end{eqnarray}

\begin{eqnarray}
\ &&\psi ((-1)^{(\widehat{\mathbf{a}})};0^{(\widehat{\mathbf{c}})})=\sqrt{%
\frac 38}\sin \theta ^{\prime }\sin ^2\theta \sin ^2\frac{\theta ^{\prime }}%
2e^{i2(\varphi -\varphi ^{\prime })}  \nonumber \\
&&\ \ \ \ \ \ \ +\sqrt{\frac 32}(3\cos ^2\frac{\theta ^{\prime }}2-\sin ^2%
\frac{\theta ^{\prime }}2)\sin \theta \cos \theta \sin ^2\frac{\theta
^{\prime }}2e^{i(\varphi -\varphi ^{\prime })}  \nonumber \\
&&\ \ \ \ \ \ \ +\sqrt{\frac 32}(3\sin ^2\frac{\theta ^{\prime }}2-\cos ^2%
\frac{\theta ^{\prime }}2)\sin \theta \cos \theta \cos ^2\frac{\theta
^{\prime }}2e^{-i(\varphi -\varphi ^{\prime })}  \nonumber \\
&&\ \ \ \ \ \ \ +\sqrt{\frac 38}(2\cos ^2\theta -\sin ^2\theta )\sin \theta
^{\prime }\cos \theta ^{\prime }  \nonumber \\
&&\ \ +\sqrt{\frac 38}\sin \theta ^{\prime }\sin ^2\theta \cos ^2\frac{%
\theta ^{\prime }}2e^{-i2(\varphi -\varphi ^{\prime })},  \label{fi59}
\end{eqnarray}
\begin{eqnarray}
\ &&\psi ((-1)^{(\widehat{\mathbf{a}})};(-1)^{(\widehat{\mathbf{c}})})=\sin
\theta ^{\prime }\sin \theta \sin ^2\frac{\theta ^{\prime }}2\sin ^2\frac
\theta 2e^{i2(\varphi -\varphi ^{\prime })}  \nonumber \\
&&\ \ \ \ \ \ \ \ +(3\cos ^2\frac{\theta ^{\prime }}2-\sin ^2\frac{\theta
^{\prime }}2)(3\cos ^2\frac \theta 2-\sin ^2\frac \theta 2)\sin ^2\frac{%
\theta ^{\prime }}2\sin ^2\frac \theta 2e^{i(\varphi -\varphi ^{\prime })} 
\nonumber \\
&&\ \ \ \ \ \ \ \ +(3\sin ^2\frac{\theta ^{\prime }}2-\cos ^2\frac{\theta
^{\prime }}2)(3\sin ^2\frac \theta 2-\cos ^2\frac \theta 2)\cos ^2\frac{%
\theta ^{\prime }}2\cos ^2\frac \theta 2e^{-i(\varphi -\varphi ^{\prime })} 
\nonumber \\
&&\ \ \ \ \ \ \ \ +\frac 32\sin \theta \sin \theta ^{\prime }\cos \theta
\cos \theta ^{\prime }+\sin \theta ^{\prime }\sin \theta \cos ^2\frac{\theta
^{\prime }}2\cos ^2\frac \theta 2e^{-i2(\varphi -\varphi ^{\prime })},
\label{si60}
\end{eqnarray}
and 
\begin{eqnarray}
\ &&\psi ((-1)^{(\widehat{\mathbf{a}})};(-2)^{(\widehat{\mathbf{c}})})=\sin
\theta ^{\prime }\sin ^2\frac{\theta ^{\prime }}2\sin ^4\frac \theta
2e^{i2(\varphi -\varphi ^{\prime })}  \nonumber \\
&&\ \ \ \ \ \ +(3\cos ^2\frac{\theta ^{\prime }}2-\sin ^2\frac{\theta
^{\prime }}2)\sin \theta \sin ^2\frac{\theta ^{\prime }}2\sin ^2\frac \theta
2e^{i(\varphi -\varphi ^{\prime })}  \nonumber \\
&&\ \ \ \ \ \ +\frac 34\sin ^2\theta \sin \theta ^{\prime }\cos \theta
^{\prime }-\sin \theta ^{\prime }\cos ^2\frac{\theta ^{\prime }}2\cos
^4\frac \theta 2e^{-i2(\varphi -\varphi ^{\prime })}  \nonumber \\
&&\ \ \ \ \ \ -(3\sin ^2\frac{\theta ^{\prime }}2-\cos ^2\frac{\theta
^{\prime }}2)\sin \theta \cos ^2\frac{\theta ^{\prime }}2\cos ^2\frac \theta
2e^{-i(\varphi -\varphi ^{\prime })},  \label{si61}
\end{eqnarray}
For the eigenvalue $-2\hbar ,$ the eigenvector is 
\begin{equation}
\lbrack \psi _{-2}]=\left( 
\begin{array}{c}
\psi ((-2)^{(\widehat{\mathbf{a}})};2^{(\widehat{\mathbf{c}})}) \\ 
\psi ((-2)^{(\widehat{\mathbf{a}})};1^{(\widehat{\mathbf{c}})}) \\ 
\psi ((-2)^{(\widehat{\mathbf{a}})};0^{(\widehat{\mathbf{c}})}) \\ 
\psi ((-2)^{(\widehat{\mathbf{a}})};(-1)^{(\widehat{\mathbf{c}})}) \\ 
\psi ((-2)^{(\widehat{\mathbf{a}})};(-2)^{(\widehat{\mathbf{c}})})
\end{array}
\right)  \label{si62}
\end{equation}
so that the probability amplitudes are 
\begin{eqnarray}
\ &&\psi ((-2)^{(\widehat{\mathbf{a}})};2^{(\widehat{\mathbf{c}})})=\cos
^4\frac \theta 2\sin ^4\frac{\theta ^{\prime }}2e^{i2(\varphi -\varphi
^{\prime })}  \nonumber \\
&&\ \ -\sin \theta ^{\prime }\sin \theta \sin ^2\frac{\theta ^{\prime }}%
2\cos ^2\frac \theta 2e^{i(\varphi -\varphi ^{\prime })}+\frac 38\sin
^2\theta ^{\prime }\sin ^2\theta  \nonumber \\
&&\ \ -\sin \theta ^{\prime }\sin \theta \cos ^2\frac{\theta ^{\prime }}%
2\sin ^2\frac \theta 2e^{-i(\varphi -\varphi ^{\prime })}  \nonumber \\
&&\ \ +\sin ^4\frac \theta 2\cos ^4\frac{\theta ^{\prime }}2e^{-i2(\varphi
-\varphi ^{\prime })},  \label{si63}
\end{eqnarray}

\begin{eqnarray}
\ &&\psi ((-2)^{(\widehat{\mathbf{a}})};1^{(\widehat{\mathbf{c}})})=\sin
\theta \sin ^4\frac{\theta ^{\prime }}2\cos ^2\frac \theta 2e^{i2(\varphi
-\varphi ^{\prime })}  \nonumber \\
&&\ -(3\sin ^2\frac \theta 2-\cos ^2\frac \theta 2)\sin \theta ^{\prime
}\sin ^2\frac{\theta ^{\prime }}2\cos ^2\frac \theta 2e^{i(\varphi -\varphi
^{\prime })}  \nonumber \\
&&\ -\frac 34\sin ^2\theta ^{\prime }\sin \theta \cos \theta -\sin \theta
\cos ^4\frac{\theta ^{\prime }}2\sin ^2\frac \theta 2e^{-i2(\varphi -\varphi
^{\prime })}  \nonumber \\
&&\ +(3\cos ^2\frac \theta 2-\sin ^2\frac \theta 2)\sin \theta ^{\prime
}\cos ^2\frac{\theta ^{\prime }}2\sin ^2\frac \theta 2e^{-i(\varphi -\varphi
^{\prime })},  \label{si64}
\end{eqnarray}

\begin{eqnarray}
\ &&\ \psi ((-2)^{(\widehat{\mathbf{a}})};0^{(\widehat{\mathbf{c}})})=\sqrt{%
\frac 38}\sin ^2\theta \sin ^4\frac{\theta ^{\prime }}2e^{i2(\varphi
-\varphi ^{\prime })}  \nonumber \\
&&\ \ \ +\sqrt{\frac 32}\sin \theta \sin \theta ^{\prime }\cos \theta \sin ^2%
\frac{\theta ^{\prime }}2e^{i(\varphi -\varphi ^{\prime })}+\sqrt{\frac 38}%
\sin ^2\theta \cos ^4\frac{\theta ^{\prime }}2e^{-i2(\varphi -\varphi
^{\prime })}  \nonumber \\
&&\ \ \ +\sqrt{\frac 3{32}}(2\cos ^2\theta -\sin ^2\theta )\sin ^2\theta
^{\prime }  \nonumber \\
&&\ \ \ -\sqrt{\frac 32}\sin \theta \sin \theta ^{\prime }\cos \theta \cos ^2%
\frac{\theta ^{\prime }}2e^{-i(\varphi -\varphi ^{\prime })},  \label{si65}
\end{eqnarray}
\begin{eqnarray}
\ &&\psi ((-2)^{(\widehat{\mathbf{a}})};(-1)^{(\widehat{\mathbf{c}})})=\sin
\theta \sin ^4\frac{\theta ^{\prime }}2\sin ^2\frac \theta 2e^{i2(\varphi
-\varphi ^{\prime })}  \nonumber \\
&&\ \ \ \ +(3\cos ^2\frac \theta 2-\sin ^2\frac \theta 2)\sin \theta
^{\prime }\sin ^2\frac{\theta ^{\prime }}2\sin ^2\frac \theta 2e^{i(\varphi
-\varphi ^{\prime })}  \nonumber \\
&&\ \ \ \ +\frac 34\sin ^2\theta ^{\prime }\sin \theta \cos \theta -\sin
\theta \cos ^4\frac{\theta ^{\prime }}2\cos ^2\frac \theta 2e^{-i2(\varphi
-\varphi ^{\prime })}  \nonumber \\
&&\ \ \ \ -(3\sin ^2\frac \theta 2-\cos ^2\frac \theta 2)\sin \theta
^{\prime }\cos ^2\frac{\theta ^{\prime }}2\cos ^2\frac \theta 2e^{-i(\varphi
-\varphi ^{\prime })},  \label{si66}
\end{eqnarray}
and 
\begin{eqnarray}
\ &&\psi ((-2)^{(\widehat{\mathbf{a}})};(-2)^{(\widehat{\mathbf{c}})})=\cos
^4\frac \theta 2\cos ^4\frac{\theta ^{\prime }}2e^{i2(\varphi -\varphi
^{\prime })}  \nonumber \\
&&\ \ \ \ +\sin \theta ^{\prime }\sin \theta \sin ^2\frac{\theta ^{\prime }}%
2\sin ^2\frac \theta 2e^{i(\varphi -\varphi ^{\prime })}+\frac 38\sin
^2\theta ^{\prime }\sin ^2\theta  \nonumber \\
&&\ \ \ \ +\sin \theta ^{\prime }\sin \theta \cos ^2\frac{\theta ^{\prime }}%
2\cos ^2\frac \theta 2e^{-i(\varphi -\varphi ^{\prime })}  \nonumber \\
&&\ \ \ \ +\cos ^4\frac \theta 2\cos ^4\frac{\theta ^{\prime }}%
2e^{-i2(\varphi -\varphi ^{\prime })}.  \label{si69}
\end{eqnarray}

\section{Discussion and Conclusion}

In this paper, we have presented the generalized probability amplitudes for
the case spin 2. The associated probabilities are obtained
straightforwardly, if tediously, as the square moduli of these probability
amplitudes. As we have emphasized, the approach to quantum mechanics that
allows us to derive and interpret these quantities leads to a deeper
understanding of quantum theory and will no doubt find application in
future. One application of the probability amplitudes for spin 2 is in the
derivation of generalized spherical harmonics; we have done this already for 
$l=1$[10]. The case $l=2$ has already been treated and will shortly be the
subject of another article. Another is in the treatment of angular momentum
addition. We have shown in previous papers how to generalize spin and/or
angular momentum addition[11,12], and this requires the generalized
probability amplitudes for the spin system being treated.

As demonstrated in our work on spin 1/2 and spin 1, generalized probability
amplitudes imply generalized forms of the spin operators. The form Eq. (\ref
{fo4}), which is considered in the literature to be the most generalized
form of the operator for the $z$ component is not so in reality. The true
generalized form will be presented in a future paper.

\section{References}

1. Mweene H. V., ''Derivation of Spin Vectors and Operators From First
Principles'', quant-ph/9905012

2. Mweene H. V., ''Generalized Spin-1/2 Operators and Their Eigenvectors'',
quant-ph/9906002

3. Mweene H. V., ''Alternative Forms of Generalized Vectors and Operators
for Spin 1/2'', quant-ph/9907031

4. Mweene H. V., ''Spin Description and Calculations in the Land\'e
Interpretation of Quantum Mechanics'', quant-ph/9907033

5. Mweene H. V., ''Vectors and Operators for Spin 1 Derived From First
Principles'', quant-ph/9906043

6. Land\'e A., ''From Dualism To Unity in Quantum Physics'', Cambridge
University Press, 1960.

7. Land\'e A., ''New Foundations of Quantum Mechanics'', Cambridge
University Press, 1965.

8. Land\'e A., ''Foundations of Quantum Theory,'' Yale University Press,
1955.

9. Land\'e A., ''Quantum Mechanics in a New Key,'' Exposition Press, 1973.

10. Mweene H. V., ''Generalized Spherical Harmonics'', quant-ph/0211135

11. Mweene H. V., ''New Treatment of Systems of Compounded Angular
Momentum'', quant-ph/9907082.

12. Mweene H. V., ''Derivation of Standard Treatment of Spin Addition From
Probability Amplitudes'', quant-ph/0003056

\end{document}